\documentclass[conference]{IEEEtran}
\pdfoutput=1
\usepackage{amsmath,epsfig,amssymb,verbatim,amsopn,subfigure,color}
\usepackage{cite,xspace}
\usepackage{array,algorithm,algorithmic}
\usepackage{bbm}
\usepackage{array}
\usepackage{multirow}
\usepackage{multicol}
\usepackage{rotating}
\usepackage{dsfont,float}
\usepackage{stfloats}
\usepackage{enumerate,makecell}
\usepackage{caption}
\usepackage[bottom]{footmisc}

\graphicspath{{figs/},{Figures/}}

\newcommand{\MeijerG}[7]{G \begin{smallmatrix} #1 & #2 \\ #3 & #4 \end{smallmatrix} \left( \begin{smallmatrix} #5 \\ #6 \end{smallmatrix} \middle\vert #7 \right) }

\newcommand{\Tr}[1]{{\color{black}{#1}}}
\newcommand{\Tb}[1]{{\color{black}{#1}}}

\newtheorem{remark}{Remark}

\newtheorem{proposition}{Proposition}

\graphicspath{{figs/},{Figures/}}

\begin{document}
\title{A Retrodirective Wireless Power Transfer Scheme for Ambient Backscatter Systems}
\author{\IEEEauthorblockN{Sahar Idrees\IEEEauthorrefmark{1}, Xiangyun Zhou\IEEEauthorrefmark{1}, Salman Durrani\IEEEauthorrefmark{1}, and Dusit~Niyato\IEEEauthorrefmark{2}}
\IEEEauthorblockA{\IEEEauthorrefmark{1}Research School of Electrical, Energy and Materials Engineering, Australian National University, Canberra, Australia.\\
Emails: \{sahar.idrees, xiangyun.zhou, salman.durrani\}@anu.edu.au.\\
\IEEEauthorrefmark{2}School of Computer Science and Engineering,
Nanyang Technological University, 50 Nanyang Ave, Singapore 639798.\\
Email: dniyato@ntu.edu.sg.}\\
}
\maketitle
\vspace{-1.5cm}
\begin{abstract}
— One of the key challenges of the Internet of Things (IoT) is to sustainably power the large number of IoT devices in real-time. \Tr{In this paper, we consider a wireless power transfer (WPT) scenario between an energy transmitter (ET) capable of retrodirective WPT and an energy receiver (ER) capable of ambient backscatter in the presence of an ambient source (AS).} The ER requests WPT by backscattering signals from an AS towards the ET, which then retrodirectively beamforms an energy signal towards the ER. To remove the inherent direct-link ambient interference, we propose a scheme of ambient backscatter training. Specifically, the ER varies the reflection coefficient multiple times while backscattering each ambient symbol according to a certain pattern called the training sequence, whose design criterion we also present. To evaluate the system performance, we derive an analytical expression for the average harvested power at the ER. Our numerical results show that with the proposed scheme, the ER harvests tens of $\mu$W of power, without any CSI estimation or active transmission from the ER, which is a significant improvement for low-power and low-cost ambient backscatter devices.
\end{abstract}

\section{Introduction}\label{Intro}
The advent and implementation of the Internet of Things (IoT) has transformed the objectives and energy requirements of the devices that constitute it. With a plethora of devices requiring pervasive and sustained connectivity, backscatter communication~\cite{cXu-2018mag} and efficient wireless power transfer (WPT) are important areas of investigation~\cite{huang2015cutting,jayakody2017wireless}. In this regard, energy beamforming is a favourable solution to increase end-to-end power efficiency~\cite{Zhang-2018survey,Alsaba2018}. \\
\indent The existing beamforming approaches either require channel state information (CSI) estimation at the ET~\cite{yZeng2015a} or the ER~\cite{gYang2014,hSon2014} or energy feedback from the ER~\cite{jXu-2014}. These schemes result in an increase in either the complexity of the ER or feedback overhead which is undesirable for IoT systems. \Tr{An alternative low complexity WPT scheme is retrodirective WPT~\cite{Lee-2018,Krikidis-2018}. The idea of retrodirective transmission involves a multi-antenna array that can transmit a beamformed signal in the source direction of an incoming wave by conjugating its phase~\cite{Miyamoto-2002}. Thus with no prior knowledge of the source direction, high link gain can be achieved while avoiding channel estimation. However, such a system always needs an interrogating signal from the prospective ER.} The work in~\cite{Lee-2018} uses active transmission at the ER to request WPT. This was avoided in~\cite{Krikidis-2018} by employing monostatic backscatter, i.e., the ET emanating a continuous un-modulated wave, which was then backscattered by the ER to request retrodirective WPT.\\
\indent Ambient backscatter systems have been gaining interest for potential employment in IoT systems due to their fundamental principle of using the existing ambient signals instead of active RF transmission. An inherent issue in ambient backscatter communication systems is the direct-link interference from the RF ambient source to the backscatter device. This is because the backscatter signal suffers inevitable attenuation and is much weaker than the original ambient signal, which has a pervasive and stronger presence. A variety of approaches have been proposed in the literature to remove or reduce this interference~\cite{Kimionis-2014,Wang-2016,Qian-2017,Yang-2018air,kellogg2016passive,bharadia2015backfi}. Some of these approaches that treat direct-link interference as a component of the background noise do not work so well since the backscatter signal is already much weaker than the ambient signal~\cite{Kimionis-2014,Wang-2016,Qian-2017}. Other schemes involve general signal processing techniques~\cite{kellogg2016passive,Yang-2018air} or backscatter specific solutions such as frequency shifting~\cite{bharadia2015backfi}.\\
\indent To the best of our knowledge, a scheme which manages WPT from the ET to the ER without channel estimation at the ET and while using low-power ambient backscatter at the ER has not been presented in the literature. We consider \Tr{an ET equipped with a phased array capable of retrodirective WPT }and an ER capable of backscattering ambient signals. One of the chief signal processing challenges of this system is then to recover the weak backscattered signal in the presence of strong direct-link ambient interference. In this context, our major \textit{contributions} are:\begin{itemize}
\item We consider an \textit{ambient backscatter training scheme} of varying the backscatter coefficient at the ER. This in effect multiplies the backscattered signal with a training sequence of $+1$ and $-1$ pulses, termed as `chips'. We analytically model the system and propose the design of this training sequence (i.e., the pattern of varying the reflection coefficient), to completely eliminate the direct-link ambient interference.
\item We show that when the ambient symbol duration is known, the ambient interference is cancelled as long as there are equal number of $+1$ and $-1$ chips over one ambient symbol. We further show that the number of chips or equivalently the switching rate does not matter in this case. Hence, we can use the slowest switching rate, i.e., we can switch the backscatter coefficient only twice per ambient symbol period.
\item We derive a closed-form expression for the average harvested power at the ER. Our results show that the average harvested power at the ER increases significantly by employing ambient backscatter training.
\end{itemize}
\section{System model}\label{sys_model}
\begin{figure}
\centering
\includegraphics[width=0.48  \textwidth]{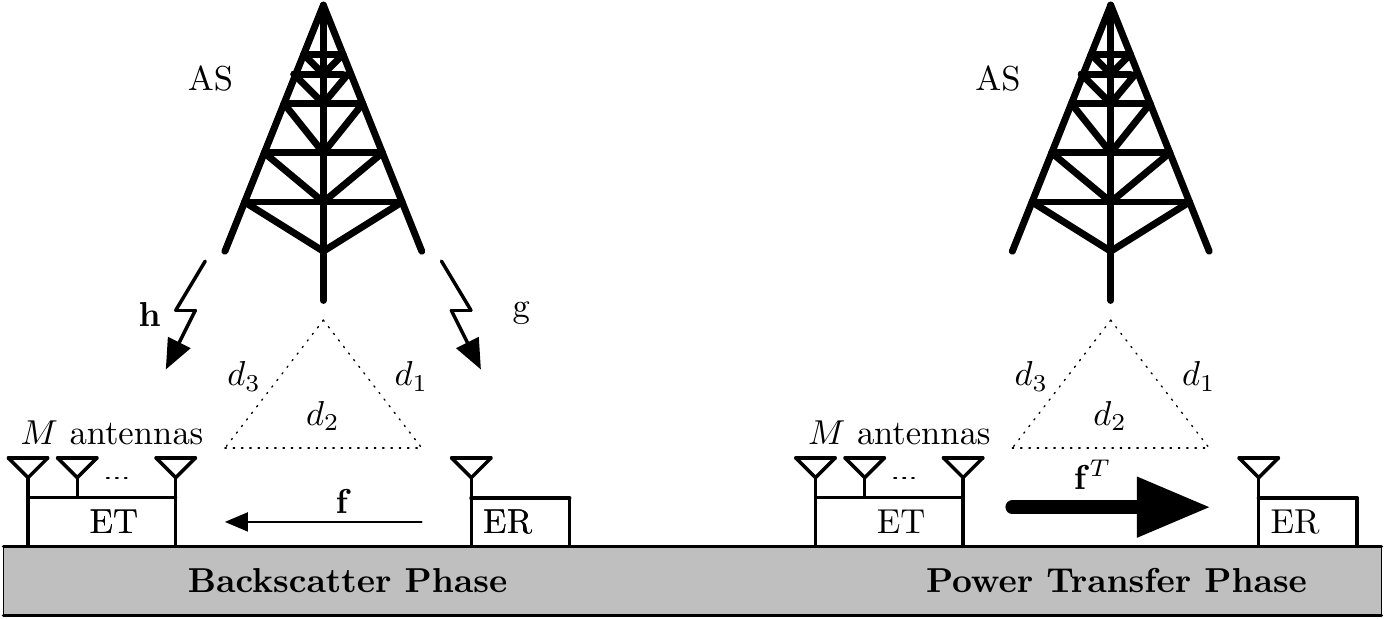}
        \caption{Illustration of the system model.}
        \label{smodel}
\end{figure}

\begin{figure}
\centering
\includegraphics[width=0.48  \textwidth]{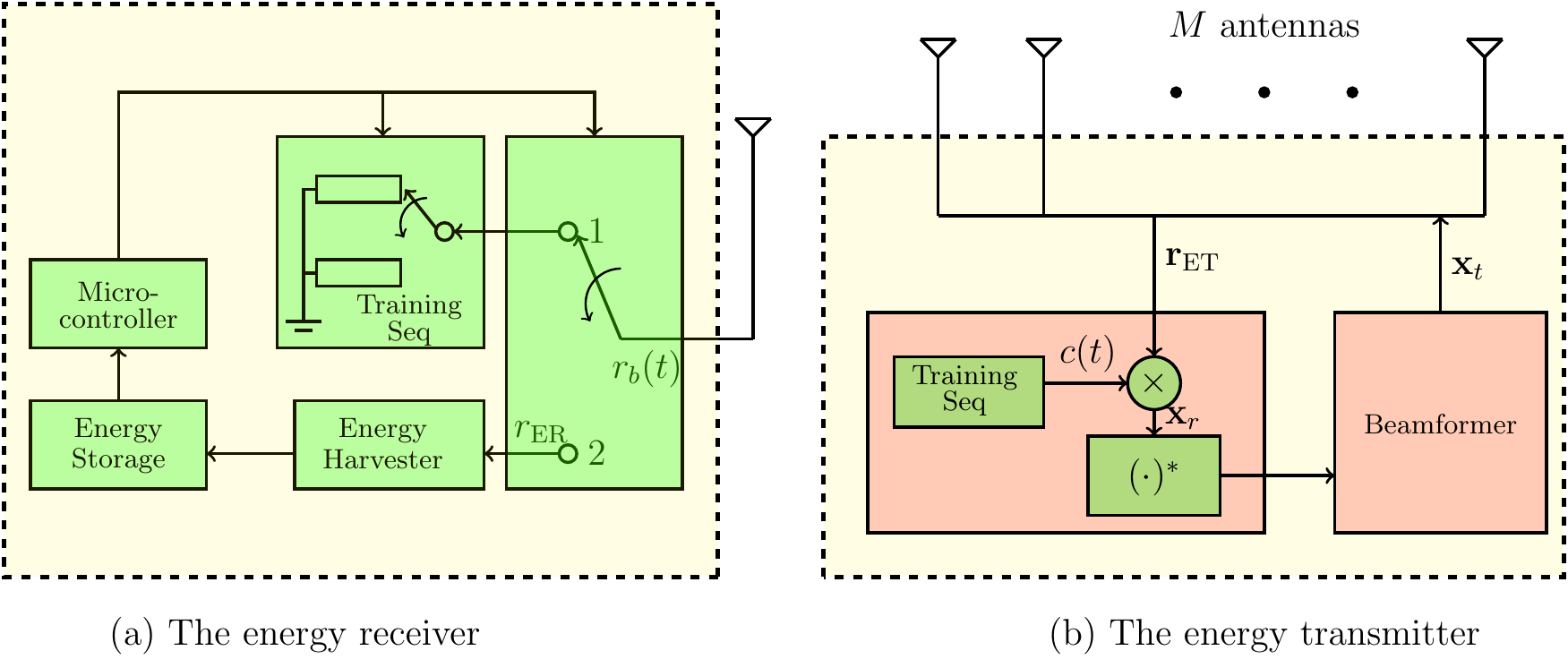}
        \caption{Block diagram of the energy transmitter and receiver.}
        \label{bdiag}
\end{figure}

\indent We study WPT in an ambient backscatter scenario including an ambient source (AS), an energy transmitter (ET) and an energy receiver (ER), as shown in Fig.~\ref{smodel}. The ET and ER both receive the signal broadcasted from the AS.\\
\indent The ER is a device capable of backscatter transmissions. It is assumed to be composed of a single antenna element, a micro-controller, a variable impedance, an energy harvester and an ideal energy storage element (e.g., a supercapacitor). \Tr{The ET is connected to a power grid and transmits with a fixed power $P_t$ using a phased array with $M$ antennas, where $M$ is large. This phased array performs retrodirective WPT. Specifically, it conjugates the phase of the incoming signal and then uses it to control the phase and amplitude of the energy signal it sends out~\cite{Lee-2018,Krikidis-2018}. In this work we have used a phased array working on the principle of retrodirective transmission as opposed to traditional `reflection type' retrodirective arrays~\cite{Pon-1964,Vanatta-1960}.} The block diagram of the ER and the ET is illustrated in Fig.~\ref{bdiag}a and~\ref{bdiag}b respectively.
\indent We assume the channel to be composed of large-scale path loss and statistically independent small-scale Rayleigh fading. We denote the distances between AS $\,\to\,$ ER, ER $\,\to\,$ ET and AS $\,\to\,$ ET by $d_1$, $d_2$ and $d_3$ respectively. Thus, large-scale attenuation is modelled as $ \gamma_i = k_0 (d_i/d_0)^{-\alpha} $ where $k_0$ is the constant attenuation for path-loss at a reference distance of $d_0$ and $i \in \{1,2,3\}$.\\
\indent The ER $\,\to\,$ ET, AS $\,\to\,$ ET and AS $\,\to\,$ ER  Rayleigh fading channel coefficients, denoted by $\mathbf{f}$, $\mathbf{h}$ and $g$ respectively, are modeled as frequency non-selective and quasi-static. \Tr{Consequently, $g$ is a circularly symmetric complex Gaussian random variable with zero mean and unit variance. Similarly, $\mathbf{f}$ and $\mathbf{h}$ are also uncorrelated circularly symmetric complex Gaussian random vectors, i.e., $\mathbf{f} = [f_1, . . . , f_M]^T  \sim  \mathcal {CN}  (0,\boldsymbol{I}_M$) and $\mathbf{h} = [h_1, . . . , h_M]^T \sim \mathcal{CN} (0,\boldsymbol{I}_M$), where $\boldsymbol{I}_M$ is the unit matrix.}\\
\indent The fading channel coefficients are assumed to be constant for a block of one backscatter phase and one power transfer phase, i.e., for $T_b + T_p$ seconds and independent and identically distributed from one $T_b + T_p$ block to the next. The use of such channels is consistent with the recent literature in this field~\cite{Krikidis-2018,Lee-2018}. Moreover, we assume  that the channel from ER to ET during the backscatter phase and the channel from ET to ER during the power transfer phase are reciprocal~\cite{Lee-2018,gYang2014,xChen2014,hSon2014,park2015joint,jXu-2014,Zeng-2015}. Finally, we do not need to make any channel state information (CSI) assumption as the use of retrodirective antenna at the ET circumvents the requirement of CSI.
\section{ Proposed Transmission Scheme}
\indent The wireless power transfer from the ET to the ER can be broken down into the following two phases:
\subsubsection{The Backscatter Phase}
\indent During the first phase, i.e., the backscatter phase of duration $T_b$ seconds, the switch in Fig.~\ref{bdiag}a stays in position 1. The ER requests the ET for WPT by sending a backscattered ambient signal to the ET. The backscattering is achieved by varying the antenna impedance mismatch, which affects the power of the reflected signal. Specifically, the signal backscattered from the ER towards the ET is
\begin{align}\label{b_backsc}
\ r_b(t) = \sqrt{\gamma_1} g \beta s(t),
\end{align}
where $\beta$ is the backscatter reflection coefficient and $s(t)$ is the ambient signal modelled as in~\cite{Tao-2018},
\begin{align}\label{ambient}
\ s(t) = \sqrt{P_s}\sum_{i=1}^{\infty} s_i p_{s}(t-iT_s),
\end{align}
where $s_i \sim  \mathcal {CN}  (0,1)$ and $p_{s}(t)$ is a rectangular pulse of unit amplitude and duration $T_s$. Note that the power of an ambient symbol in~\eqref{ambient} is $P_s$.
 The ET receives this backscattered signal from the ER as well as the direct signal from the ambient source and noise. In general, the backscattered signal component in the received signal at the ET is much weaker than the original ambient signal (also referred to as direct-link interference) since it suffers pathloss and attenuation twice.\\
\indent \Tr{In order to resolve this issue we perform ambient backscatter training. In this regard, we assume a BPSK-like backscatter coefficient having two possible values, i.e., $\beta = \pm 1$.}\footnote{$\beta$ can assume any pair of values $|\beta| \leq 1$. However, for simplicity we assume that $|\beta| = 1$.} The ambient backscatter training means that while backscattering the ambient signal, the ER switches the backscatter coefficient between $+1$ and $-1$ $N_c$ times, according to a pre-defined training sequence,\Tr{ known to the ET.} This switching is done at a rate of $\frac{1}{T_c}$, where $T_c$ is the duration for which the backscatter coefficient maintains a certain value. This is effectively equivalent to multiplying the backscattered signal with a training signal $c(t)$ of $N_c$ short duration pulses, which we call `chips', each of duration $T_c$ and of amplitudes $+1$ and $-1$. Thus, at a given time instant $t$, the backscattered signal from the ER is
\begin{align}\label{back_sig}
\ r_b(t) = \sqrt{ \gamma_1} g c(t)s(t),
\end{align}
where $c(t)$ is the training signal with $N_c$ pulses and chip duration $T_c$. It can be modelled as
\begin{align}\label{chip_seq}
\ c(t) = \sum_{n=0}^{N_c-1} c_n p_c(t-nT_c),
\end{align}
where $c_n$ is the $n$-th chip ($+1$ or $-1$) of the training signal and $p_c(t)$ is a rectangular pulse of unit amplitude and duration $T_c$. This training signal applied at the ER is quite similar to the Direct Sequence Spread Spectrum (DSSS)~\cite{goldsmith-2005}.\\
\indent The received signal at the ET is given by
\begin{align}\label{rec_sig}
 \mathbf{r}_\textrm{ET}(t) = \sqrt{\gamma_1 \gamma_2} g \mathbf{f} c(t)s(t)  + \sqrt{\gamma_3} \mathbf{h} s(t) + \mathbf{n}(t) ,
\end{align}
where $\mathbf{n}(t)\sim \mathcal{CN} (0,{\sigma_n}^2 \boldsymbol{I}_M)$ is the AWGN. Note that $\mathbf{r}_{ET}(t)$ is a composite signal with three components, i.e., the backscattered signal from the ER, the ambient signal from the AS and the AWGN.
The ET correlates this composite signal with the known training signal with \Tb{perfect frame synchronization. The case of imperfect synchronization has been discussed in the journal version of this paper~\cite{Idrees-2019}}. Thus the signal at the output of the correlator is given as,
\begin{align}\label{despread}
\mathbf{x}_{r}&= \frac{1}{\ N_cT_c} \int_{0}^{\ N_cT_c} \mathbf{r}_\textrm{ET} (t) c(t) dt\nonumber\\
&= \underbrace{\frac{1}{\ N_cT_c}\int \limits_{0}^{N_cT_c} \sqrt{\gamma_1\gamma_2} g \mathbf{f} c(t)s(t)c(t)  dt}_{\mathbf{x}_{s}} \nonumber\\
& + \underbrace{\frac{1}{\ N_cT_c}\int \limits_{0}^{N_cT_c}\sqrt{\gamma_3} \mathbf{h} s(t)c(t) dt}_{\mathbf{x}_{i}}+ \underbrace{\frac{1}{\ N_cT_c}\int \limits_{0}^{N_cT_c} \mathbf{n}(t) c(t)dt}_ {\mathbf{\widetilde{n}}},
\end{align} where $\mathbf{x}_{s}$ and $\mathbf{x}_{i}$ are desired signal and undesired ambient (i.e., interference) components at the output of the correlator.
\indent Substituting the value of $c(t)$ from~\eqref{chip_seq}, we get $\mathbf{x}_{s}$ as
\begin{align}
\mathbf{x}_{s}&= \frac{\sqrt{\gamma_1 \gamma_2} g \mathbf{f}}{\ N_cT_c} \int \limits_{0}^{N_cT_c}\sum_{i=1}^{N_s} s_i p_{s}(t-iT_s)\nonumber\\
& \sum_{n=0}^{N_c-1} \sum_{m=0}^{N_c-1} c_n p_c(t-nT_c) c_m p_c(t-mT_c) dt, \label{xs1a}
\end{align}
\indent The integration in~\eqref{xs1a} is being performed for the product of two aligned rectangular pulses $p_c(t)$ and $p_s(t)$ where $T_s \geq T_c$ and the duration of integration is $N_cT_c$, so we get
\begin{subequations}
\begin{alignat}{4}
 \mathbf{x}_{s}&= \frac{\sqrt{\gamma_1 \gamma_2 P_s}g \mathbf{f}}{N_cT_c} \sum_{i=1}^{N_s} s_i\sum_{n=\frac{N_c}{N_s}(i-1)}^{\frac{N_c}{N_s}i-1}c_n^2 \int \limits_{nT_c}^{(n+1)T_c}  {p_c}^2(t-nT_c) dt,\label{xs1b}\\
&= \sqrt{\gamma_1 \gamma_2 P_s} \frac{ g \mathbf{f}}{\ N_cT_c } \sum_{i=1}^{N_s} s_i  \sum_{n=\frac{N_c}{N_s}(i-1)+1}^{\frac{N_c}{N_s}i}T_c, \label{xs1c}\\
&= \sqrt{\gamma_1 \gamma_2 P_s} \frac{ g \mathbf{f}}{\ N_c } \frac{N_c}{N_s} \sum_{i=1}^{N_s} s_i = \sqrt{\gamma_1 \gamma_2 P_s} \frac{ g \mathbf{f}}{N_s} \sum_{i=1}^{N_s} s_i, \label{xs1e}
\end{alignat}
\end{subequations}
\noindent  Also,~\eqref{xs1c} follows from the fact that $c_n^2 = 1$ and $\int_{nT_c}^{(n+1)T_c}  {p_c}^2(t-nT_c) dt = T_c$.~\eqref{xs1e} can be written since $\sum_{n=\frac{N_c}{N_s}(i-1)}^{\frac{N_c}{N_s}i-1} = \frac{N_c}{N_s}$ for any given $i$.\\
\indent Similarly we get $\mathbf{x}_{i}$  as
\begin{subequations}
\begin{alignat}{3}
 \mathbf{x}_{i} &=  \frac{\sqrt{\gamma_3} \mathbf{h}}{ N_cT_c} \int \limits_{0}^{N_cT_c} \sqrt{P_s} \sum_{i=1}^{N_s} s_i p_{s}(t-iT_s) \sum_{n=1}^{N_c} c_n p_c(t-nT_c)dt, \label{xi1a}\\
 &= \sqrt{\gamma_3 P_s} \frac{ \mathbf{h}}{N_cT_c} \sum_{i=1}^{N_s}s_i \sum_{n=\frac{N_c}{N_s}(i-1)}^{\frac{N_c}{N_s}i-1} c_n  \int \limits_{nT_c}^{(n+1)T_c} p_c(t-nT_c)dt, \label{xi1b}\\
&= \sqrt{\gamma_3 P_s} \frac{ \mathbf{h}}{N_cT_c} \sum_{i=1}^{N_s} s_i \sum_{n=\frac{N_c}{N_s}(i-1)}^{\frac{N_c}{N_s}i-1} c_n T_c, \label{xi1c}\\
&= \sqrt{\gamma_3 P_s} \frac{ \mathbf{h}}{N_c} \sum_{i=1}^{N_s} s_i \sum_{n=\frac{N_c}{N_s}(i-1)}^{\frac{N_c}{N_s}i-1} c_n,  \label{xi1d}
\end{alignat}
\end{subequations}
\noindent where again the integration in~\eqref{xi1a} becomes the summation in~\eqref{xi1c} as mentioned above.\\
\subsubsection{The Power Transfer Phase}
During the second phase, i.e., the power transfer phase of duration $T_p$ seconds, the ET provides retrodirective wireless power transfer to the ER. Specifically, \Tr{the large phased array at the ET} conjugates the phase of the signal at the output of the correlator in~\eqref{despread} and hence beamforms towards the ER. Thus, the signal transmitted by the ET is a single-tone sinusoidal waveform subject to the maximum total transmit power $P_t$ at the ET. It is given as,
\begin{align}\label{xt}
\mathbf{x}_t &= \sqrt{P_t} \frac{(\mathbf{x}_{r})^*}{\left\|\mathbf{x}_{r}\right\|},
\end{align}
where $\left\|\mathbf{x}_{r}\right\| = \sqrt{{\mathbf{x}_{r}}^T \mathbf{x}_{r}}$. Note that in~\eqref{xt}, we have dropped the time index $t$ because the baseband signal $x_t$ does not vary with time.
The switch in the ER in Fig.~\ref{bdiag}a now moves to position 2. Consequently, the ER stops backscattering and only harvests energy from the incoming beam with energy harvesting efficiency $\zeta$. This energy is then stored in the energy storage device in the ER. Analytically, the signal received by the ER in the power transfer phase is given by
\begin{align}\label{rER}
{r}_{\textrm{ER}} &= \sqrt{\gamma_2}\mathbf{f}^T \mathbf{x}_t, \nonumber\\
&= \sqrt{\gamma_2 P_t} \frac{\left(\mathbf{f}^T  {\mathbf{x}_{s}}^*+ \mathbf{f}^T  {\mathbf{x}_{i}}^*+ \mathbf{f}^T {\mathbf{\widetilde{n}}}^*\right)}{\left\|\mathbf{x}_{s}+ \mathbf{x}_{i} + \mathbf{\widetilde{n}}\right\|},
\end{align}
where $\mathbf{x}_{s}$ is given in~\eqref{xs1e}, $\mathbf{x}_{i}$ is given in~\eqref{xi1d} and $ \mathbf{\widetilde{n}}\sim \mathcal{CN} (0,\frac{{\sigma_n}^2}{N_cT_c}\boldsymbol{I}_M)$ is the noise at the output of the matched filter. Note that the receiver noise at the ER is not included in~\eqref{rER} because it is irrelevant to energy harvesting.\\
\indent In this work, we use the \textit{average harvested power} at the ER, $\overline{Q}$, as the figure of merit. It is defined as
\begin {align}\label{q_def}
\overline{Q} = E[Q] = E[\zeta |{r}_{\textrm{ER}}|^2],
\end{align}
where $Q$ is the instantaneous harvested power and $\zeta$ is the RF-to-DC current energy conversion efficiency. In this work we assume unit time in the power transfer phase. Hence, we use the terms energy and power interchangeably.
\section{Training Sequence Design for the Removal of Direct-Link Ambient Interference}\label{detseq}
As mentioned earlier, in addition to the backscattered ambient signal from the ER, the ET also receives the original ambient signal which is orders of magnitude stronger than its backscattered version from the ER. This is due to the fact that the backscatter signal suffers attenuation twice, i.e., in going from AS to ER and then from ER to ET. As a result, it is considerably weakened and the signal received at the ET during the backscatter phase is predominantly composed of the ambient component.\\
\indent This signal processing problem of recovering the weak backscatter signal in the presence of a much stronger unwanted ambient signal is quite similar to the signal recovery problem in the direct sequence spread spectrum (DSSS). Taking inspiration from that, we apply a training sequence to the backscattered signal at the ER during the backscatter phase. However, in the interest of avoiding power leakage towards the AS, we aim to eliminate the direct-link interference from the AS at the ET altogether. To this end, we propose the design of a deterministic training sequence instead of using a PN sequence. The \textit{Design Criterion} for such a training sequence is presented below. We assume that the number of ambient symbols in the backscatter phase is $N_s$, i.e., $T_b = N_sT_s = N_cT_c$.\\
\indent \underline{\textit{Design Criterion:}} For the system model considered in Section~\ref{sys_model}, the ambient component can be eliminated at the output of the correlator in the ET if for each ambient symbol that is backscattered from the ER during the backscatter phase, the number of $+1$ and $-1$ chips is equal, i.e., $N_{+1} = N_{-1}$ and $N_{+1} + N_{-1} = \frac{N_c}{N_s}$, where $N_{+1}$ and $N_{-1}$ are the number of positive and negative chips respectively that are multiplied per symbol of the ambient source. This means that the backscatter coefficient is switched between $+1$ and $-1$ an even number of times, i.e., $N_c = 2kN_s$ where $k$ is a positive integer.
\begin{IEEEproof}
We justify the above design criterion as follows:
$c(t)$ is a deterministic sequence of equal number of $+1$ and $-1$ chips. Any sequence with equal number of $+1$ and $-1$ chips applied to each ambient symbol while backscattering, does the job. We can see from~\eqref{xs1e} that, the desired backscattered component at the output of the correlator $\mathbf{x}_\textrm{s}$ does not depend on the attributes of the training sequence, i.e., how the backscatter coefficient is changed. Therefore, it remains the same as in the previous case. However, with our proposed training sequence satisfying the \textit{design criterion},~\eqref{xi1d} becomes
\begin{align}\label{xival}
\mathbf{x}_i &= \sqrt{\gamma_3 P_s} \frac{ \mathbf{h}}{N_c} \sum_{i=1}^{N_s} s_i \sum_{n= \frac{N_c}{N_s}(i-1)}^{\frac{N_c}{N_s}i - 1} c_n, \nonumber\\
& =   \sqrt{\gamma_3 P_s} \frac{ \mathbf{h}}{N_c} \sum_{i=1}^{N_s} s_i \left[ (+1)N_{+1} + (-1)N_{-1} \right] =0
\end{align}
since $N_{+1} = N_{-1}$.
Thus, the ambient component at the output of the correlator cancels out.
\end{IEEEproof}
\begin{remark}
  The \textit{design criterion} is generic, i.e., any sequence that satisfies the two properties can serve the purpose. Moreover, we have seen that once the ambient component is removed, having a greater number of chips does not affect the harvested energy. Therefore, taking into account the practical implementation, it is best to have the minimum number of chips per ambient symbol period, i.e., $k = 1$ and $N_c = 2N_s$ or $T_c = \frac{T_s}{2}$. This means that we can switch the backscatter coefficient only twice per ambient symbol, i.e., for each ambient symbol that is backscattered, the backscatter coefficient is kept $+1$ for half of the ambient symbol duration and $-1$ for the other half.
\end{remark}

\indent Using the proposed sequence in the \textit{design criterion}, we find the average harvested power at the ER, which is presented in the proposition below.
\begin{proposition}\label{p3}
For the system model considered in Section~\ref{sys_model} and employing the backscatter training scheme proposed in the \textit{design criterion}, the instantaneous harvested power at the ER at the end of the power transfer phase when the number of antennas at the ET $M\,\to\, \infty$, is given by
\begin{align}\label{q_instp}
Q \approx \zeta\gamma_2 P_t \left(\frac{\gamma_1 \gamma_2 |g|^2 \mu (M+1) + \frac{\sigma_n^2 N_s}{T_s P_s}} {\gamma_1 \gamma_2 |g|^2 \mu  + \frac{\sigma_n^2 N_s}{T_s P_s}}\right).
\end{align}
From~\eqref{q_instp} we can obtain the average harvested power at the ER, given by
\begin{align}\label{q_avg}
 \overline{Q} & \approx \frac{\zeta P_t \sigma_n^2}{T_s P_s \gamma_1} (M + 1) \MeijerG {3}{1}{1}{3}{-1}{-1,0,0}{\frac{\sigma_n^2}{P_s T_s \gamma_1 \gamma_2}}\nonumber \\
  &+ \frac{\zeta P_t \sigma_n^2}{T_s P_s \gamma_1} \MeijerG {3}{1}{1}{3}{0}{0,0,0}{\frac{\sigma_n^2}{P_s T_s \gamma_1 \gamma_2}},
\end{align}
where $\mu$ is as defined in~\eqref{mu} and $\MeijerG{m}{n}{p}{q}{a_1,\ldots,a_p}{b_1,\ldots,b_q}{z}$ is the MeijerG function~\cite{gradshteyn2007}.
\end{proposition}
\begin{IEEEproof}
See Appendix.
\end{IEEEproof}
\section{Results}
\begin{figure}
\centering
\includegraphics[width=0.5  \textwidth]{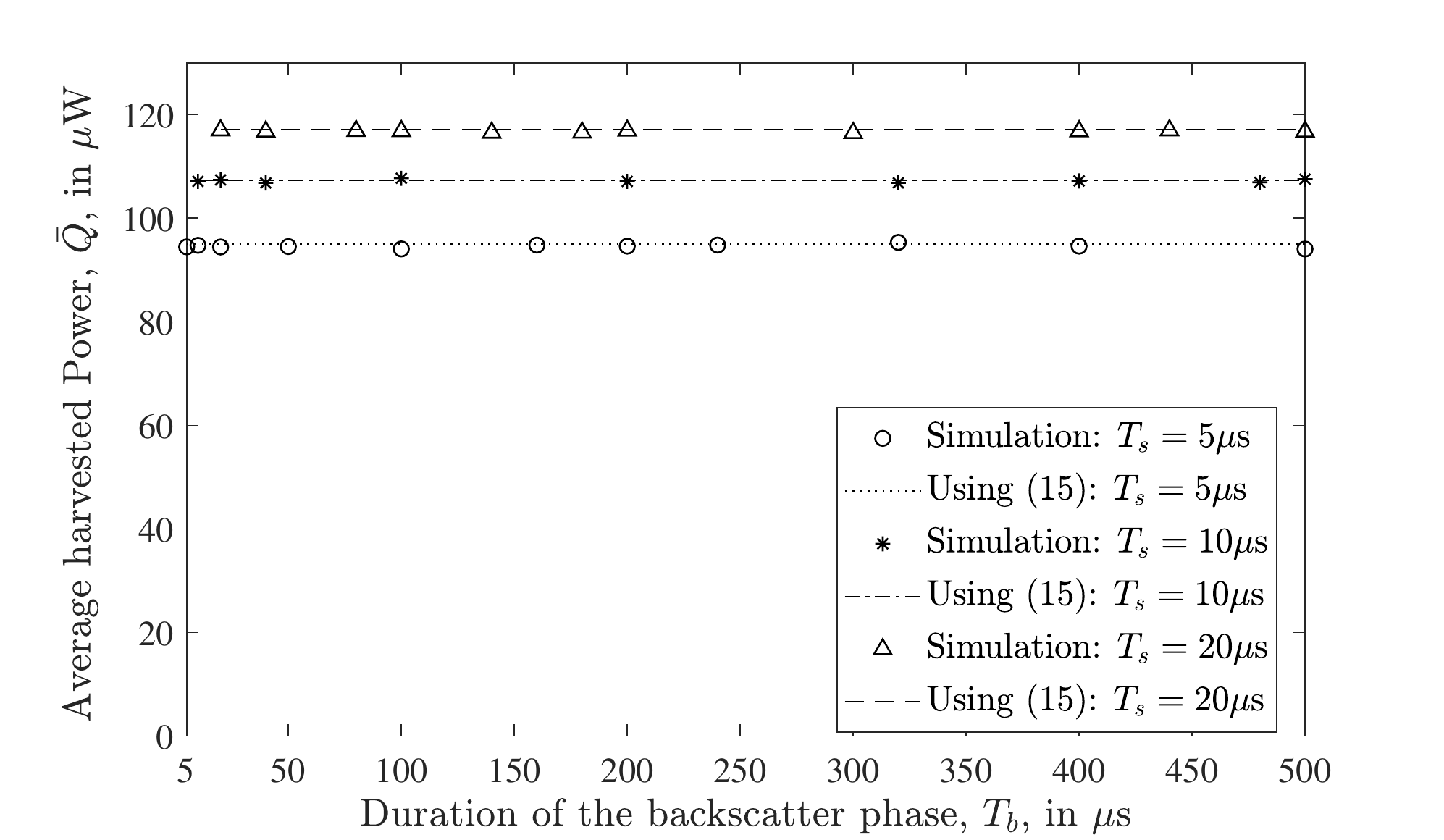}
        \caption{Average harvested power at the ER with the proposed sequence plotted against the duration of the backscatter phase, $T_b$.}
        \label{eh2}
\end{figure}

\begin{figure}
\centering
  \includegraphics[width=.5\textwidth]{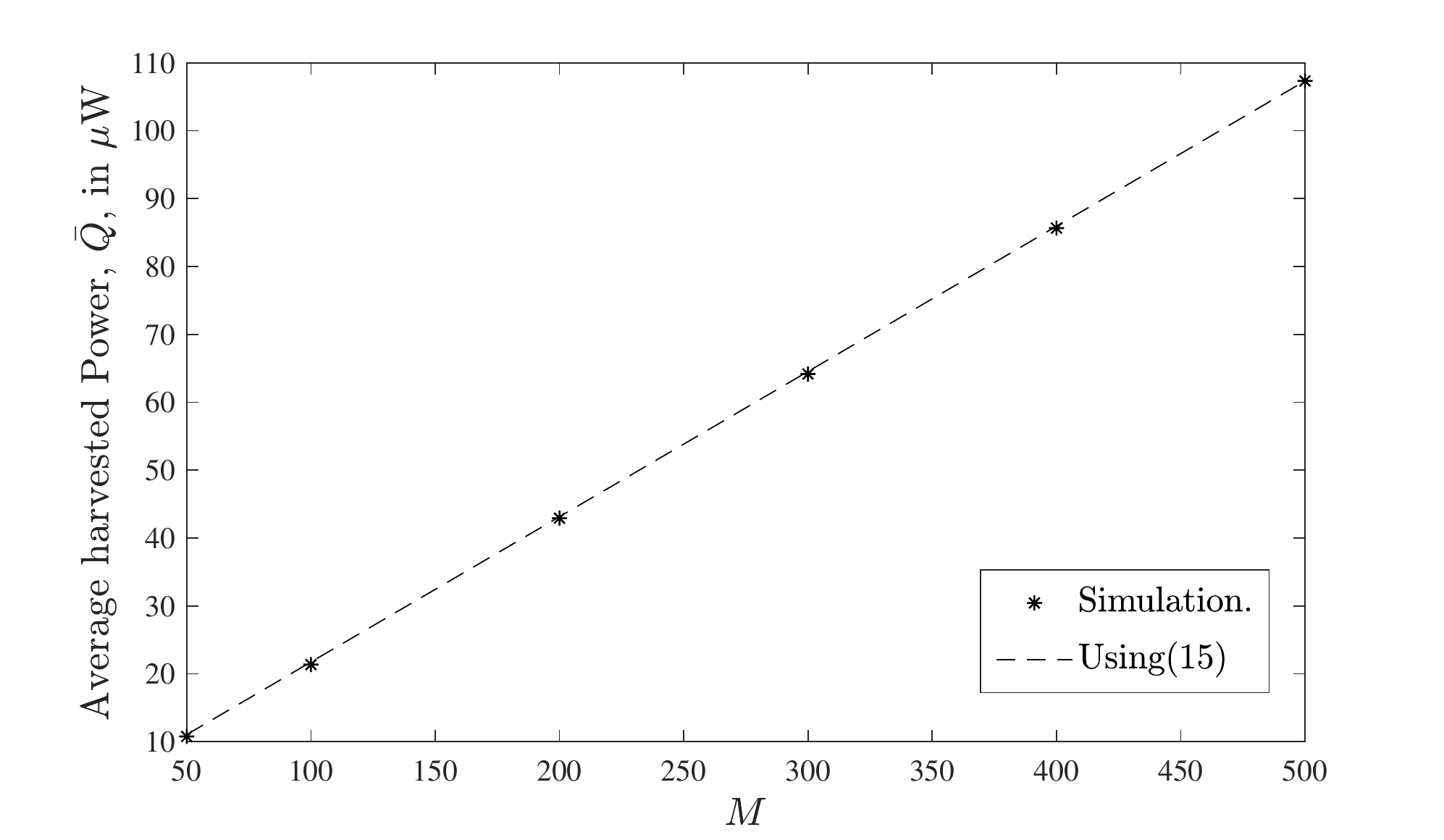}
  \caption{Average harvested power, $\bar{Q}$, plotted against number of antennas at the ET $M$.}
  \label{M_plots}
\end{figure}

In this section, we present numerical results evaluating the system performance under our proposed design. For the results presented in this section, the values of different system parameters are given in Table~\ref{simT}.
 The choice of $T_c = 500$ ns ensures that the multipath delay spread is negligible~\cite{bharadia2015backfi}. All the results presented in this section are averaged over $10^4$ Monte Carlo simulation trials.\\
 \begin{table}[t]
\centering
\caption{System Parameter Values.}\label{simT}
\begin{tabular}{|c|l|c|c|} \hline
 No. & Parameter & Symbol & Value \\ \hline%
  1. & Reference Distance  & $d_0$ & 1m \\
  2. & Distance between the AS and the ER  & $d_1$ & 10m \\
  3. & Distance between the ER and the ET &$d_2$&  20m\\
  4. & Distance between the AS and the ET & $d_3$& 18m \\
  5. & Constant attenuation for path loss & $k_0$ & 0.001\\
  6. & Path-loss exponent & $\alpha$& 2.5\\
  7. &  Transmit power of the AS & $P_s$& 1W \\
  8. &  Transmit power of the ET & $P_t$ & 1W\\
  9. & Variance of AWGN & $\sigma_n^2$& $10^{-18}$\\
  10. & Number of antennas at the ET & $M$ & 500   \\
  11. & Chip duration &${T_c}$& 500 ns  \\
 12. & Energy harvesting efficiency & ${\zeta}$& 0.5\\
 \hline
\end{tabular}
\end{table}
\indent \underline{\textit{Need for Ambient Backscatter Training:}} For the purpose of comparison, we calculated the average harvested power at the ER versus the duration of the backscatter phase, i.e., $T_b$ without this particular scheme of ambient backscatter training i.e., varying the reflection coefficient for $T_s = 5~\mu$s. The harvested power in this case is only a fraction of a $\mu$W, i.e., $0.28 \mu$W, which is not very useful. As mentioned before, this is due to the majority of power leaking towards the AS due to the relative strength of direct-link ambient interference.\\
\indent \underline{\textit{Impact of $T_b$ and $T_s$:}} Next, using~\eqref{q_avg}, the average harvested power at the ER is plotted in Fig.~\ref{eh2} for three different values of ambient symbol duration, i.e., $T_s = 5~\mu$s, $T_s = 10~\mu$s and $T_s = 20~\mu$s with the proposed scheme employed. Numerous features of the proposed scheme are evident from Fig.~\ref{eh2}.\\
\indent Firstly, it can be observed that the energy harvested at the ER increases significantly. This is due to the fact that the proposed scheme completely eliminates the ambient component. As a result, the retrodirective beamformer forms a focused beam directed back at the ER alone, with no energy leaking towards the AS.\\
 \indent Secondly, the harvested power at the ER does not change with the increase in backscatter training duration, but \textit{stays constant as long as the ambient symbol duration $T_s$ stays constant.} This is confirmed by the result given in~\eqref{q_avg} as the average harvested energy is independent of the duration of the backscatter phase $T_b$. Specifically, when the system is designed with a fixed value of $T_c$, then for different values of $N_s$ and hence $T_b$, the average harvested power at the ER now stays at $95~\mu$W for $T_s = 5~\mu$s, $108.3~\mu$W for $T_s = 10~\mu$s and $116.4~\mu$W for $T_s = 20~\mu$s.\\
 \indent Thirdly, with the ambient component removed, the average harvested power \textit{depends largely on the duration of the ambient symbol $T_s$,} as is evident from the plot with the average harvested power having a larger value for $T_s = 20~\mu$s, compared to $T_s = 5~\mu$s \\
  \indent Lastly, it can also be inferred from the plot that for a fixed ambient source, the average harvested power in this case is \textit{independent of the number of chips $N_c$.} Actually, for a fixed chip duration, the number of chips also increases with the increased backscatter period $T_b$ and as we can see from Fig.~\ref{eh2}, the average harvested power stays constant for the increased values of the backscatter period.\\
\indent \underline{\textit{Impact of $M$:}} Fig.~\ref{M_plots} plots the average harvested power against $M$, the number of antennas at the ET. It can be observed that there is a good agreement again between the results obtained by simulation and by using~\eqref{q_avg} for practical values of $M$.
 \section{Conclusions}
In this work we have presented a wireless power transfer scheme to energize an ER using \Tr{retrodirective WPT} at the ET and ambient backscatter at the ER. To deal with the direct-link ambient interference, we proposed the approach of backscatter training, i.e., the pattern of varying the reflection coefficient at the ER to completely eliminate the strong direct-link ambient interference. We showed that when the ambient symbol duration is known, the switching rate does not matter and we can switch the backscatter coefficient only twice per ambient symbol period. Future work can consider this problem in a multiple ER and/or multiple ET scenario.
\begin{appendices}
\section{Proof of Proposition 1}\label{a}
We derive the formula for instantaneous energy harvested at the ER during the power transfer phase as given in~\eqref{q_avg}. Substituting the values of $\mathbf{x}_s$ and $\mathbf{x}_i$ from ~\eqref{xs1e} and~\eqref{xival} into~\eqref{rER}, we get
\begin{align}\label{rERapprox}
{r}_{\textrm{ER}} &= \sqrt{\gamma_2 P_t} \frac{(\frac{\sqrt{\gamma_1 \gamma_2 P_s} g^*}{N_s} \sum_{i=1}^{N_s} s_i^* {\textbf{f}}^T \mathbf{f}^* + \mathbf{f}^T {\mathbf{n}}^* )}{\left\|\frac{\sqrt{\gamma_1 \gamma_2 P_s} g^*}{N_s} \sum_{i=1}^{N_s} s_i {\textbf{f}} + {\mathbf{n}}\right\|}.
\end{align}
\indent Since ${\mathbf{f}}^T \mathbf{f}^* = {\mathbf{f}}^H \mathbf{f}$ and ${\mathbf{f}}^T \mathbf{h}^* = {\mathbf{f}}^H \mathbf{h}$,~\eqref{rERapprox} simplifies to
\begin{align}\label{rERapprox2}
{r}_{\textrm{ER}} &= \sqrt{\gamma_2 P_t} \frac{(\frac{\sqrt{\gamma_1 \gamma_2 P_s} g}{N_s} \sum_{i=1}^{N_s} s_i^* {\textbf{f}}^H \textbf{f} + \mathbf{f}^H {\mathbf{n}} )}{\left\|\frac{\sqrt{\gamma_1 \gamma_2 P_s} g}{N_s} \sum_{i=1}^{N_s} s_i {\textbf{f}}+ {\mathbf{n}}\right\|}.
\end{align}
\indent Substituting~\eqref{rERapprox2} in~\eqref{q_def}, we get
\begin{align}\label{qapprox}
Q &= \zeta |r_{\textrm{ER}}|^2 \nonumber\\
&= \zeta \left({\frac{\frac{\gamma_1 {\gamma_2}^2 P_s P_t |g|^2}{N_s^2} \left|\sum_{i=1}^{N_s} s_i\right|^2 {\left\| \mathbf{f} \right\|}^4 + \gamma_2 P_t {\left\|\mathbf{f}^H {\mathbf{n}}\right\|}^2 } {\frac{{\gamma_1 \gamma_2 P_s} |g|^2}{N_s^2} \left|\sum_{i=1}^{N_s} s_i^*\right|^2  {\left\| \mathbf{f} \right\|}^2 +  \|{\mathbf{n}}\|^2 }}\right).\nonumber
\end{align}
Let
\begin{equation}\label{mu}
\mu = \left|\sum_{i=1}^{N_s} s_i\right|^2 = \left|\sum_{i=1}^{N_s} s_i^*\right|^2
\end{equation}
and using the asymptotic massive MIMO expressions~\cite{Lim-2015}, i.e. $\frac{1}{M} {\left\|\mathbf{f}_i\right\|}^4 \,\to\, M +1 $, $\frac{1}{M} {\left\|\mathbf{f}_i\right\|}^2 \,\to\, 1 $, $\frac{1}{M} {\left\|{\mathbf{f}_k}^H  \mathbf{f}_i \right\|}^2 \,\to\, 1 $, $\frac{1}{M} {\mathbf{f}_k}^H  \mathbf{f}_i  \,\to\, 0 $ (for $k \neq i$),  $\frac{1}{M} {\mathbf{f}_k}^H  \widetilde {\mathbf{n}}  \,\to\, 0 $, $\frac{1}{M} {\left\|{\mathbf{f}_i}^H\widetilde {\mathbf{n}} \right\|}^2 \,\to\, \frac{{\sigma_n}^2}{NT_c} $ and $\frac{1}{M} {\left\|{\widetilde {\mathbf{n}}} \right\|}^2 \,\to\, \frac{{\sigma_n}^2}{NT_c} $, and since $N_c T_c = N_sT_s$, we get the result in~\eqref{q_instp} reproduced below
\begin{align}\label{qa}
Q \approx \zeta\gamma_2 P_t \left(\frac{\gamma_1 {\gamma_2}  |g|^2 \mu (M+1) + \frac{\sigma_n^2 N_s}{T_s P_s}} {\gamma_1 \gamma_2 |g|^2 \mu + \frac{\sigma_n^2 N_s}{T_s P_s}}\right).
\end{align}
\indent The average harvested power at the ER is then given by,
\begin{equation}\label{exp_q}
\overline{Q} = E[Q]= \int_{0}^{\infty} \zeta\gamma_2 P_t \left(\frac{\gamma_1 \gamma_2 z (M+1) + \frac{\sigma_n^2 N_s}{T_s P_s}} {\gamma_1 \gamma_2 z  + \frac{\sigma_n^2 N_s}{T_s P_s}}\right) f_Z(z) dz,
\end{equation}
where $z = |g|^2 \mu$ and $|g|^2$ and $\mu$ are independent of each other. Since $|g|^2 \sim$ Exp($1$) and $\mu \sim$ Exp($\frac{1}{N_s}$), $f_Z(z) = \frac{2}{N_s} K_0 (2\sqrt{\frac{z}{N_s}})$ for $z \geq 0$, where $K_0$ is the bessel function of the first kind. Using Mathematica,~\eqref{exp_q} can be expressed as~\eqref{q_avg}.
\end{appendices}

\end{document}